\documentclass[12pts]{article}
\usepackage[textwidth=15cm]{geometry}
\usepackage{graphicx}	
\expandafter\let\csname equation*\endcsname\relax
\expandafter\let\csname endequation*\endcsname\relax
\usepackage{amsmath,cases}	
\usepackage{notoccite}
\usepackage{soul}
\usepackage{xcolor}

\begin{document}

\title{Anisotropic ultracompact Schwarzschild star\\
 by gravitational decoupling}
\author{J. Ovalle$^{1,2}$, C. Posada$^1$ and Z. Stuchl\'ik$^1$\\
$^{1}$ {\small Institute of Physics and Research Centre of Theoretical Physics and Astrophysics}\\
{\small Faculty of Philosophy and Science}\\
{\small Silesian University in Opava}\\
{\small Bezru\v{c}ovo n\'{a}m. 13, CZ-74601 Opava, Czech Republic}\\
$^{2}$ {\small Departamento de F\'isica, Universidad Sim\'on Bol\'ivar}\\
{\small AP 89000, Caracas 1080A, Venezuela}}

\maketitle
\vspace{10pt}

%
\begin{abstract}
We employ the minimal geometric deformation approach to gravitational decoupling (MGD-decoupling) in order to generate an exact anisotropic and non-uniform version of the ultracompact Schwarzschild star, or ‘gravastar', proposed by Mazur and Mottola. This new system represents an ultracompact configuration of radius $R_S=2\,{\cal M}$ whose interior metric can be matched smoothly to a conformally deformed Schwarzschild exterior. Remarkably, the model satisfies some of the basic requirements to describe a stable stellar model, such as a positive density everywhere and decreasing monotonously from the centre, as well as a non-uniform and monotonic pressure.
\end{abstract}
\vspace{2pc}
\section{Introduction}
%

Black holes are one of the most active areas of gravitational physics, mainly due to the fact that they constitute ideal laboratories to test general relativity in the strong field regime. However, confronting theoretical predictions with observations is an arduous and complicated task. A formidable step in this direction is the recent direct observation of black holes shadows \cite{Akiyama:2019cqa,Akiyama:2019bqs}, as well as the observation of black holes through the detection of gravitational waves \cite{Abbott:2016blz,Abbott:2017oio,Abbott:2017gyy}, which opens a new and promising era for gravitational physics. 


Despite the strong evidence in favour of their existence, black holes still present some paradoxes which have not been solved satisfactorily \cite{Wald:1999vt}. This has motivated some authors to propose some alternatives, or black-hole ‘mimickers' (see \cite{cardoso2019} for a recent review), which can be compact enough to generate a shadow as the one recently observed~\cite{Akiyama:2019cqa,Akiyama:2019bqs}. One of these mimickers is the gravastar model of Mazur and Mottola \cite{mazur2001,mazur2004}. In the original gravastar scenario, a collapsing star suffers a phase transition at, or close to, where the horizon would have been formed. As a result, the interior region is replaced by a patch of de Sitter spacetime with negative pressure $p=-\rho$. This interior region is matched to the exterior Schwarzschild spacetime, through a shell of stiff matter $p=\rho$. The gravastar provides a final state for gravitational collapse, with no central singularity nor an event horizon, therefore avoiding the issues that these notions entail in classical black holes. 

Despite the fact that the gravastar has been widely studied in the literature (see e.g. \cite{Cattoen:2005he,Lobo:2006xt,Chirenti:2007mk,Chirenti:2008pf,Chirenti:2016hzd,Cardoso:2007az,Pani:2009ss,Horvat:2011ar,Cardoso:2014sna,sakai2014,Pani:2015tga,Volkel:2017ofl,Nakao:2018knn} and references therein), its mechanism of formation remains as the biggest challenge for this model. Mazur and Mottola \cite{Mazur:2015kia} unveiled a connection between the gravastar and the interior Schwarzschild solution \cite{schwarzschild1916b} or `Schwarzschild star', characterised by a constant energy density and isotropic pressure. Although the Schwarzschild star is an ideal case which might not be fully attained in a realistic physical scenario (see however \cite{misner1973}), nevertheless it provides a simple analytical solution to Einstein's equations which allows further analysis \cite{Stuchlik:2008xe,Stuchlik:2016xiq,Boehmer:2003uz}.  

Mazur and Mottola \cite{Mazur:2015kia} investigated the Schwarzschild star in the ultracompact regime beyond the Buchdahl limit $R=(9/4)M$ \cite{buchdahl1959}, and showed that the limiting configuration, when the radius of the star approaches the Schwarzschild radius $R\to R_{S}$, becomes one with a regular interior of constant negative pressure $p=-\rho$ determined by a patch of modified de Sitter spacetime. This sphere of ‘dark energy' is bounded by a boundary layer of anisotropic stresses, located at the Schwarzschild radius, endowed with certain surface tension. By Birkhoff's theorem, the exterior remains the exterior Schwarzschild spacetime.  The ultracompact Schwarzschild star has zero entropy and temperature, so mirroring the main characteristics of the gravastar proposed in \cite{mazur2001,mazur2004}. 
    
The ultracompact Schwarzschild star was extended to slow rotation in \cite{Posada:2016qpz}.  A result of particular interest is that the moment of inertia and mass quadrupole moment, or I-Q relations, are in agreement with the corresponding Kerr values \cite{Urbanec:2013fs}. More recently in \cite{Camilo:2018goy} it was shown that the Schwarzschild star is stable against radial perturbations. As an extension of these results, Konoplya et. al. \cite{Konoplya:2019nzp} showed that the ultracompact Schwarzschild star is stable against non-radial (axial) gravitational perturbations. Moreover they showed that the $l>1$ perturbations are indistinguishable from those of Schwarzschild black holes. The results above asserts the viability of the ultracompact Schwarschild star as a legitimate black hole mimicker. Gravitational perturbations of more general polytropic spheres were studied in \cite{Stuchlik:2017qiz}.

Inspired in the results above, it is relevant to extend the Schwarzschild star to a more general scenario, while preserving its fundamental properties. However, extending a known solution to a more complex situation can be an overwhelming task, given the complexity of Einstein's field equations~\cite{Stephani}. Fortunately, the so-called method of gravitational decoupling by Minimal Geometric Deformation (MGD-decoupling, henceforth)~\cite{Ovalle:2017fgl,Ovalle:2019qyi}, which has been widely used recently~\cite{Ovalle:2017wqi,Gabbanelli:2018bhs,Ovalle:2018umz,Sharif:2018toc,Contreras:2018gzd,Contreras:2018vph,Morales:2018nmq,Heras:2018cpz,Panotopoulos:2018law,Sharif:2018tiz,Contreras:2019iwm,Maurya:2019wsk,Contreras:2019fbk,Ovalle:2018ans,Sharif:2018khl},
 has proved to be a powerful method to extend known solutions into more complex frameworks.
 \par
 The original version of the MGD approach was developed in Refs.~\cite{Ovalle:2007bn,Ovalle:2009xk} in the context of extra-dimensional gravity~\cite{Randall:1999ee,Randall:1999vf}, and it was eventually extended to study black hole solutions in
 Refs.~\cite{Casadio:2015gea,Ovalle:2015nfa} (for some earlier works on the MGD, see for instance
 Refs.~\cite{Casadio:2012pu,Ovalle:2013xla,Ovalle:2013vna,Casadio:2013uma},
 and Refs.~\cite{Ovalle:2014uwa,Casadio:2015jva,Cavalcanti:2016mbe,Casadio:2016aum,daRocha:2017cxu,daRocha:2017lqj,Fernandes-Silva:2017nec,Casadio:2017sze,Fernandes-Silva:2018abr,Fernandes-Silva:2019fez} for some recent applications). The MGD-decoupling has three main characteristics that make it particularly useful in the search
 for new solutions of Einstein's field equations, namely:
 \begin{itemize}
 	\item
 	We can extend any solution of the Einstein equations into more complex domains. For instance, we can start from a source with energy-momentum tensor $\hat T_{\mu\nu}$ for which the metric is known
 	and add the energy-momentum tensor of a second source,
 	\begin{equation}
 	\label{coupling0}
 	\hat T_{\mu\nu}
 	\rightarrow
 	T_{\mu\nu}
 	=
 	\hat T_{\mu\nu}
 	+T^{(1)}_{\mu\nu}
 	\ .
 	\end{equation}
 	We can then repeat the process with more sources $T^{(i)}_{\mu\nu}$ to extend the solution of the Einstein equations
 	associated with the gravitational source $\hat T_{\mu\nu}$ into the domain of more intricate forms of gravitational sources
 	$T_{\mu\nu}$;
 	\item
 	We can reverse the previous procedure in order to find a solution to Einstein's equations with a complex
 	energy-momentum tensor ${T}_{\mu\nu}$ by separating it into simpler components,
 	\begin{equation}
 	\label{split}
 	{T}_{\mu\nu}
 	\rightarrow
 	\hat T_{\mu\nu}+T^{(i)}_{\mu\nu}
 	\ ,
 	\end{equation} 
 	and solve Einstein's equations for each one of these components.
 	Hence, we will have as many solutions as the components in the original energy-momentum tensor ${T}_{\mu\nu}$.
 	Finally, by a simple combination of all these solutions, we will obtain the solution to the Einstein equations associated
 	with the original energy-momentum tensor ${T}_{\mu\nu}$.
 	\item
 	We can apply it to theories beyond general relativity.
 	For instance, given the modified action~\cite{Ovalle:2019qyi}
 	\begin{equation}
 	\label{ngt}
 	S_{\rm G}
 	=
 	S_{\rm EH}+S_{\rm X}
 	=
 	\int\left[\frac{R}{2\,k^2}+{\cal L}_{\rm M}+{\cal L}_{\rm X}\right]\sqrt{-g}\,d^4\,x
 	\ ,
 	\end{equation}
 	where ${\cal L}_{\rm M}$ contains all matter fields in the theory and ${\cal L}_{\rm X}$ is the Lagrangian density
 	of a new gravitational sector with an associated energy-momentum tensor
 	\begin{equation}
 	\label{ngt2}
 	\theta_{\mu\nu}
 	=
 	\frac{2}{\sqrt{-g}}\frac{\delta(\sqrt{-g}\,{\cal L}_{\rm X})}{\delta g^{\mu\nu}}
 	=
 	2\,\frac{\delta{\cal L}_{\rm X}}{\delta g^{\mu\nu}}-g_{\mu\nu}\,{\cal L}_{\rm X}
 	\ ,
 	\end{equation}
 	we can use~\eqref{coupling0} to extend all the known solutions of the Einstein-Hilbert action $S_{\rm EH}$
 	into the domain of modified gravity represented by $S_{\rm G}$.
 	This represents a straightforward way to study the consequences of extended gravity on general relativity. 
 \end{itemize}
In this paper we will apply the procedure describe in~\eqref{coupling0} to extend the Mazur-Mottola model in order to build a new ultracompact interior configuration
 with non-uniform matter density and anisotropic pressure.
 \par
 The paper is organised as follows: in Section~\ref{s2} we present Einstein's equations for a spherically symmetric stellar configuration and we discuss how to decouple two spherically symmetric and static gravitational sources
 $\{T_{\mu\nu},\,\theta_{\mu\nu}\}$, as well as
 the matching conditions at the stellar surface under the MGD-decoupling. In Section~\ref{s3} we review the constant-density interior Schwarzschild solution, or Schwarzschild star, and the negative pressure regime.  In Section~\ref{s4}, we implement the MGD-decoupling following the scheme~\eqref{coupling0} to generate the extended
 version of the Mazur-Mottola model. Finally, in Section~\ref{con} we summarise our conclusions.
 
 
%
\section{Gravitational decoupling of two sources by MGD}
\label{s2}
\setcounter{equation}{0}
Let us start from the standard Einstein field equations~\footnote{We use the metric signature $(+---)$
	and the constant $k^2=8\,\pi\,G_{\rm N}$.}
\begin{eqnarray}
\label{EinEq}
R_{\mu\nu}-\frac{1}{2}\,R\,g_{\mu\nu}
=
k^2\,T^{\rm (tot)}_{\mu\nu}
\ ,
\end{eqnarray}
where the energy-momentum tensor $T^{\rm (tot)}_{\mu\nu}$ is given by
\begin{eqnarray}
\label{emt}
T^{\rm (tot)}_{\mu\nu}
=
{T}_{\mu\nu}+\theta_{\mu\nu}
\ ,
\end{eqnarray}
where ${T}_{\mu\nu}$ and  $\theta_{\mu\nu}$ represent two generic gravitational sources. Let us recall that the Einstein tensor is divergenceless and therefore the total energy momentum tensor ${T}^{\rm (tot)}_{\mu\nu}$ must satisfy
the conservation equation
\begin{eqnarray}
\nabla_\nu\,T^{{\rm (tot)}{\mu\nu}}
=
0
\ .
\label{dT0}
\end{eqnarray}
In Schwarzschild-like coordinates, the spherically symmetric metric reads 
\begin{eqnarray}
ds^{2}
=
e^{\nu (r)}\,dt^{2}-e^{\lambda (r)}\,dr^{2}
-r^{2}\left( d\theta^{2}+\sin ^{2}\theta \,d\phi ^{2}\right)
\ ,
\label{metric}
\end{eqnarray}
where $\nu =\nu (r)$ and $\lambda =\lambda (r)$ are functions of the areal radius $r$ only, ranging from the center $r=0$ up to the stellar surface $r=R>0$. Explicitly, the field equations read
\begin{eqnarray}
\label{ec1}
&&
k^2
\left(T_0^{\,0}
+\theta_0^{\,0}
\right)
=
\strut\displaystyle\frac 1{r^2}
-e^{-\lambda }\left( \frac1{r^2}-\frac{\lambda'}r\right)\ ,
\\
&&
\label{ec2}
k^2
\strut\displaystyle
\left(T_1^{\,1}+\theta_1^{\,1}\right)
=
\frac 1{r^2}-e^{-\lambda }\left( \frac 1{r^2}+\frac{\nu'}r\right)\ ,
\\
&&
\label{ec3}
k^2
\strut\displaystyle
\left(T_2^{\,2}+\theta_2^{\,2}\right)
=
\frac 14e^{-\lambda }\left[ -2\,\nu''-\nu'^2+\lambda'\,\nu'
-2\,\frac{\nu'-\lambda'}r\right]
\ ,
\end{eqnarray}
while the conservation equation, which is a linear combination of \eqref{ec1}-\eqref{ec3}, yields
\begin{eqnarray}
\label{con1}
&&
\left({T}_1^{\ 1}\right)'
-
\frac{\nu'}{2}\left({T}_0^{\ 0}-{T}_1^{\ 1}\right)
-
\frac{2}{r}\left({T}_2^{\ 2}-{T}_1^{\ 1}\right)
\nonumber
\\
&&+
\left({\theta}_1^{\ 1}\right)'
-
\frac{\nu'}{2}\left({\theta}_0^{\ 0}-{\theta}_1^{\ 1}\right)
-
\frac{2}{r}\left({\theta}_2^{\ 2}-{\theta}_1^{\ 1}\right)
=
0
\ .
\end{eqnarray}
where $f'\equiv \partial_r f$. By simple inspection of \eqref{ec1}-\eqref{ec3}, we can identify an effective density 
\begin{eqnarray}
\tilde{\rho}
=
T_0^{\,0}
+\theta_0^{\,0}
\ ,
\label{efecden}
\end{eqnarray}
an effective isotropic pressure
\begin{eqnarray}
\tilde{p}_{r}
=-T
_1^{\,1}-\theta_1^{\,1}
\ ,
\label{efecprera}
\end{eqnarray}
and an effective tangential pressure
\begin{eqnarray}
\tilde{p}_{t}
=-T
_2^{\,2}-\theta_2^{\,2}
\ .
\label{efecpretan}
\end{eqnarray}
The expressions above clearly illustrate the appearance of an anisotropy inside the stellar distribution, given by 
\begin{eqnarray}
\label{anisotropy}
\Pi
\equiv
\tilde{p}_{t}-\tilde{p}_{r}.
\end{eqnarray}

Equations \eqref{ec1}-\eqref{ec3} contain five unknown functions, namely, two metric functions $\{\nu(r),\,\lambda(r)\}$ and three physical variables: the density $\tilde{\rho}(r)$, the radial pressure $\tilde{p}_r(r)$
and the tangential pressure $\tilde{p}_t(r)$. Thus these equations form an indefinite system \cite{Herrera:1979,Mak:2001eb} which requires additional information to produce any specific solution.

In order to solve the Einstein equations~\eqref{ec1}-\eqref{con1} we implement the MGD-decoupling. 
In this approach, one starts from a solution to \eqref{EinEq}  for the source $T_{\mu\nu}$
[that is \eqref{ec1}-\eqref{con1} with $\theta_{\mu\nu}=0$] such that 
the metric reads

\begin{eqnarray}\label{pfmetric}
ds^{2}=e^{\xi (r)}\,dt^{2}-e^{\mu(r)}\,dr^{2}-r^{2}\left( d\theta^{2}+\sin ^{2}\theta \,d\phi ^{2}\right)\ ,
\end{eqnarray}
where 
\begin{eqnarray}
\label{standardGR}
e^{-\mu(r)} \equiv 1-\frac{k^2}{r}\int_0^r x^2\,\rho\, dx
=
1-\frac{2\,m(r)}{r},
\end{eqnarray}
is the standard General Relativity expression containing the Misner-Sharp mass function $m=m(r)$.
Next, we turn on the second source $\theta_{\mu\nu}$ to see its effects on the first source $T_{\mu\nu}$. These effects are encoded in the geometric deformation undergone by the geometry \eqref{pfmetric}, namely
\begin{eqnarray}
\label{gd1}
\xi
&\mapsto &
\nu
=
\xi+\alpha\,g
\ ,
\\
\label{gd2}
e^{-\mu}
&\mapsto &
e^{-\lambda}
=
e^{-\mu}+\alpha\,f
\ ,
\end{eqnarray}
where $g$ and $f$ are, respectively, the deformations undergone by the temporal and radial
metric component of the geometry $\{\xi,\mu\}$.
Among all possible deformations~$\{g,f\}$, the simplest one is the so-called minimal geometric
deformation given by~$\{g=0,f=f^{*}\}$, and therefore only the radial metric component changes to
\begin{eqnarray}
\label{expectg}
e^{-\mu(r)}\mapsto\,e^{-\lambda(r)}
=
e^{-\mu(r)}+\alpha\,f^{*}(r)
\ .
\end{eqnarray}
The system~\eqref{ec1}-\eqref{con1} can be decoupled by plugging the deformation~(\ref{expectg})
into the Einstein equations~(\ref{ec1})-(\ref{ec3}).
The system is thus separated into two sets of equations:
(i) one having the standard Einstein field equations for the energy-momentum tensor $T_{\mu\nu}$,
whose metric is given by \eqref{pfmetric} with $\xi(r)=\nu(r)$,
\begin{eqnarray}
\label{ec1pf}
&&
k^2\,T_0^{\,0}
=
\strut\displaystyle\frac 1{r^2}
-e^{-\mu }\left( \frac1{r^2}-\frac{\mu'}r\right)\ ,
\\
&&
\label{ec2pf}
k^2\,T_1^{\,1}
=
\frac 1{r^2}-e^{-\mu }\left( \frac 1{r^2}+\frac{\nu'}r\right)\ ,
\\
&&
\label{ec3pf}
k^2\,T_2^{\,2}
=
\frac 14e^{-\mu }\left[ -2\,\nu''-\nu'^2+\mu'\,\nu'
-2\,\frac{\nu'-\mu'}r\right]
\ ,
\end{eqnarray}
along with the conservation equation~(\ref{dT0}) with $\theta_{\mu\nu} = 0$, namely $\nabla_\nu\,{T}^{{\mu\nu}}=0$, yielding
\begin{eqnarray}
\label{conpf}
\left({T}_1^{\ 1}\right)'
-
\frac{\nu'}{2}\left({T}_0^{\ 0}-{T}_1^{\ 1}\right)
-
\frac{2}{r}\left({T}_2^{\ 2}-{T}_1^{\ 1}\right) = 0\ ,
\end{eqnarray}
which is a linear combination of \eqref{ec1pf}-\eqref{ec3pf}; and (ii) one for the source $theta_{\mu\nu}$, which reads
\begin{eqnarray}
\label{ec1d}
&&
k^2\,\theta_0^{\,0}
=
-\strut\displaystyle\frac{\alpha\,f^{*}}{r^2}
-\frac{\alpha\,f^{*'}}{r}\ ,
\\
&&
\label{ec2d}
k^2
\strut\displaystyle
\,\theta_1^{\,1}
=- \alpha\,f^{*}\left(\frac{1}{r^2}+\frac{\nu'}{r}\right)\ ,
\\
&&
\label{ec3d}
k^2
\strut\displaystyle\,\theta_2^{\,2}
=
-\frac{\alpha\,f^{*}}{4}\left(2\nu''+\nu'^2+\frac{2\,\nu'}{r}\right)-\frac{\alpha\,f^{*'}}{4}\left(\nu'+\frac{2}{r}\right)
\ .
\end{eqnarray}
Its conservation equation  $\nabla_\nu\,\theta^{\mu\nu}=0$ explicitly reads
\begin{eqnarray}
\label{con1d}
(\theta_1^{\,\,1})'
-\strut\displaystyle\frac{\nu'}{2}(\theta_0^{\,\,0}
-\theta_1^{\,\,1})-\frac{2}{r}(\theta_2^{\,\,2}-\theta_1^{\,\,1})
=
0
\ ,
\end{eqnarray}
which is a linear combination of \eqref{ec1d}-\eqref{ec3d}. We recall that, under these conditions, there is no exchange of energy-momentum between the perfect fluid and the source $\theta_{\mu\nu}$ and therefore their interaction is purely gravitational. 
\subsection{Deformed vacuum and matching conditions at the surface}
\label{s5}
\par
Let us recall the matching conditions at the stellar surface $r=R$ between the interior geometry ($0\le r\le R$) of a self-gravitating system and the exterior $(r>R)$ spacetime. The interior is described by the generic metric~\eqref{metric}, which in terms of the MGD transformation~\eqref{expectg} reads
\begin{eqnarray}
ds^{2}
=
e^{\nu^{-}(r)}\,dt^{2}
-\left[1-\frac{2\,\tilde{m}(r)}{r}\right]^{-1}dr^2
-r^{2}\left(d\theta ^{2}+\sin {}^{2}\theta d\phi ^{2}\right)
\ ,
\label{mgdmetric}
\end{eqnarray}
where the interior mass function is given by
\begin{eqnarray}
\label{effecmass}
\tilde{m}(r)
=
m(r)-\frac{r}{2}\,\alpha\,f^{*}(r)
\ , 
\end{eqnarray} 
with the Misner-Sharp mass $m$ given by \eqref{standardGR} and $f^{*}$ the geometric deformation in \eqref{expectg}. On the other hand, the exterior spacetime will be described by the deformed Schwarzschild metric 
\begin{eqnarray}
\label{MetricSds}
ds^2=\left(1-\frac{2{\cal M}}{r}\right)dt^2-\left(1-\frac{2{\cal M}}{r}+\beta\,g^{*}(r)\right)^{-1}dr^2-d\Omega^2
\ ,
\end{eqnarray}
which determines the Schwarzschild vacuum $T^{+}_{\mu\nu}=0$ filled by a generic energy-momentum tensor $\theta^{+}_{\mu\nu}\neq\,0$. We remark that this exterior could be filled by fields contained in the source $\theta^{+}_{\mu\nu}$. The function $g^{*}(r)$ in the metric~\eqref{MetricSds} is precisely the geometric deformation for the outer Schwarzschild solution due to $\theta^{+}_{\mu\nu}$. 
Notice that the interior and exterior deformations are different, likewise as their respective parameters $\alpha$ and $\beta$.
\par
 After decoupling the Schwarzschild vacuum $T^{+}_{\mu\nu}=0$ and $\theta^{+}_{\mu\nu}\neq\,0$, the set of equations~\eqref{ec1d}-\eqref{ec3d} for the exterior $r>R$ reads
\begin{eqnarray}
\label{ec1de}
&&
k^2\,(\theta_0^{\,0})^+
=
-\strut\displaystyle\frac{\beta\,g^{*}}{r^2}
-\frac{\beta\,g^{*'}}{r}\ ,
\\
&&
\label{ec2de}
k^2
\strut\displaystyle
\,(\theta_1^{\,1})^+
=-\frac{\beta\,g^{*}}{r\,(r-2\,{\cal M})}\ ,
\\
&&
\label{ec3de}
k^2
\strut\displaystyle\,(\theta_2^{\,2})^{+}
=
\frac{{\cal M}\,(r-{\cal M})}{r^2\,(r-2\,{\cal M})^2}\,\beta\,g^{*}-\frac{(r-{\cal M})}{2\,r\,(r-2\,{\cal M})}\beta\,g^{*'}
\ ,
\end{eqnarray}
together with their respective conservation equations, which are a linear combination of \eqref{ec1de}-\eqref{ec3de}. 
\par
The metrics in \eqref{mgdmetric} and \eqref{MetricSds} must satisfy the Israel-Darmois matching
conditions~\cite{Israel:1966rt} at the surface $\Sigma$ defined by $r=R$, namely, the continuity of the first and second fundamental form. Continuity of the first fundamental form reads 
\begin{eqnarray}
\left[ ds^{2}\right] _{\Sigma }=0
\ ,
\label{match1}
\end{eqnarray}
where 
$[F]_{\Sigma }\equiv F(r\rightarrow R^{+})-F(r\rightarrow R^{-})\equiv F_{R}^{+}-F_{R}^{-}$,
for any function $F=F(r)$, which yields \begin{eqnarray}
e^{\nu ^{-}(R)}
=
1-\frac{2\,{\cal M}}{R}
\ ,
\label{ffgeneric1}
\end{eqnarray}
and
\begin{eqnarray}
1-\frac{2\,M}{R}+\alpha\,f^{*}_{R}
=
1-\frac{2\,{\cal M}}{R}+\beta\,g^{*}_{R}
\ ,
\label{ffgeneric2}
\end{eqnarray}
where $M=m(R)$, with $f^{*}_{R}$ and $g^{*}_{R}$ being the interior and exterior minimal geometric deformation evaluated at the star surface, respectively. Likewise, continuity of the second fundamental form reads
\begin{eqnarray}
\left[G_{\mu \nu }\,r^{\nu }\right]_{\Sigma }
=
0
\ ,
\label{matching1}
\end{eqnarray}
where $r_{\mu }$ is a unit radial vector.
Using \eqref{matching1} and the general Einstein equations~(\ref{EinEq}),
we then find 
\begin{eqnarray}
\left[T_{\mu \nu }^{\rm (tot)}\,r^{\nu }\right]_{\Sigma}
=
0
\ ,
\label{matching2}
\end{eqnarray}
which leads to 
\begin{eqnarray}
\left[T_1^{\,\,1}+\theta_1^{\,\,1}\right]_{\Sigma }
=
0
\ .
\label{matching3}
\end{eqnarray}
This matching condition takes the final form 
\begin{eqnarray}
(T_1^{\,\,1})^{-}_{R}+\,(\theta_1^{\,\,1})^{-}_{R}
=(\theta_1^{\,\,1})^{+}_{R}
\ .
\label{matchingf}
\end{eqnarray} 
The condition in Eq.~(\ref{matchingf}) is the general expression for the second fundamental form associated
with the Einstein equations~(\ref{EinEq}) and the Schwarzschild vacuum filled by a generic source $\theta^{+}_{\mu\nu}$, namely,  $\{T^{+}_{\mu\nu}=0,\,\theta^{+}_{\mu\nu}\,\neq\,0\}$. 

\par
By using Eqs.~(\ref{ec2d}) and~\eqref{ec2de} in the condition~(\ref{matchingf}), the second fundamental
form can be written as
\begin{eqnarray}
-k^2\,(T_1^{\,\,1})^{-}_{R}
+\alpha\,f_{R}^{*}\left(\frac{1}{R^{2}}+\frac{\nu _{R}^{\prime }}{R}\right)
=
\frac{\beta\,g_{R}^{*}}{r\,(r-2\,{\cal M})}
\ ,
\label{sfgenericf}
\end{eqnarray}
where $\nu _{R}^{\prime }\equiv \partial _{r}\nu^{-}|_{r=R}$. In terms of the effective pressure \eqref{efecprera}, we can express the condition~\eqref{sfgenericf} as
\begin{eqnarray}
\tilde{p}^-_R=\tilde{p}^+_R\ ,
\end{eqnarray}
which establishes the continuity of the effective radial pressure at the stellar surface. The expressions in \eqref{ffgeneric1}, \eqref{ffgeneric2} and \eqref{sfgenericf} are the necessary
and sufficient conditions for the matching of the interior MGD metric \eqref{mgdmetric}
to a spherically symmetric outer ‘vacuum' described by the deformed Schwarzschild
metric {\eqref{MetricSds}}.
\par
Finally, we remark an important result regarding the matching condition~(\ref{sfgenericf}):
if the outer geometry is given by the Schwarzschild metric, namely, $g^{*}(r) = 0$ in \eqref{MetricSds}, then the condition \eqref{sfgenericf} reads

\begin{eqnarray}
\tilde{p}^{-}_R=\,p_{R}+\alpha\,\frac{f_{R}^{\ast }}{k^2}
\left(\frac{1}{R^{2}}+\frac{\nu _{R}^{\prime }}{R}\right)=0
\ .
\label{pnegative}
\end{eqnarray}
Therefore the star will be in equilibrium in a vacuum only if the effective radial pressure at the surface vanishes. In particular, if the inner geometric deformation $f^*(r<R)$ is positive and weakens the gravitational field [see \eqref{effecmass}], an outer Schwarzschild can be only compatible with a non-vanishing inner $\theta_{\mu\nu}$ if the isotropic stellar matter has $p_{R}<0$ at the surface of the star. This could be interpreted as regular matter with a solid crust~\cite{Ovalle:2014uwa}. If we want to avoid having a solid-crust and keep the standard condition $p_{R}=0$, we must impose
that the anisostropic effects on the radial pressure vanish at $r=R$. This can be achieved if we assume that $(\theta_1^{\,\,1})^{-}_{R}\sim\,p_{R}$ in \eqref{matchingf}, which leads to a vanishing inner deformation $f^*_R=0$. 
\section{‘Gravastar' as the ultracompact Schwarzschild star} \label{s3}

In this section we briefly discuss the Schwarzschild interior solution, or Schwarzschild star, corresponding to a uniform-density spherical star, and the natural emergence of the interior region with negative pressure (see \cite{Mazur:2015kia,Posada:2016qpz} for a more detailed discussion). We start by considering a spherically symmetric spacetime given by

\begin{eqnarray}
ds^2 = +e^{\nu(r)}dt^2-e^{\lambda(r)}dr^2-r^2(d\theta^2+\sin^2\theta d\phi^2).
\end{eqnarray}

The interior Schwarzschild solution can be written in the following form \cite{misner1973}

\begin{numcases}{e^{\nu(r)} = }
\frac{1}{4}\left(3\sqrt{1-H^2R^2}-\sqrt{1-H^2r^2}\right)^2, & $r < R$, \label{interiorf} \\ 
\left(1-\frac{2M}{r}\right), & $r > R$,
\end{numcases}

\noindent and

\begin{numcases} {e^{-\lambda(r)} =}
1-H^2r^2, & $r < R$, \label{interiorh} \\ 
\left(1-\frac{2M}{r}\right), & $r > R$,  
\end{numcases}

\noindent where $M$ is the mass and $R$ is the radius of the star. Here we have defined

\begin{eqnarray}
H^2\equiv\frac{8\pi\rho_{0}}{3}=\frac{R_{S}}{R^3}=\frac{2M}{R^3},
\end{eqnarray}
  
\noindent where $\rho_{0}$ is the energy density, which is constant, and $R_{S}$ is the Schwarzschild radius. The pressure is determined by

\begin{eqnarray}\label{pressure}
p(r) = \rho_{0}\left(\frac{\sqrt{1-H^2r^2} - \sqrt{1-H^2R^2}}{3\sqrt{1-H^2R^2}-\sqrt{1-H^2r^2}}\right).
\end{eqnarray}

\noindent The Buchdahl condition $R/M>9/4$, guarantees that the pressure is positive and finite everywhere inside the star \cite{buchdahl1959}. However, it can be observed that the pressure \eqref{pressure} is regular, except at some radius $R_{0}$ given by

\begin{eqnarray}\label{pole}
R_{0}=3R\sqrt{1-\frac{8}{9}\frac{R}{R_{S}}}<R.
\end{eqnarray}

\noindent Moreover, as it can be observed from \eqref{interiorf} and \eqref{pressure}, the pressure diverges exactly at the same point where $g_{tt}=0$. Nevertheless, if we consider the Schwarzschild star in the regime $2<R/M<9/4$, an interesting behaviour can be observed; the pole in the pressure \eqref{pole} moves out from the center of the star to a finite surface $R_{0}$ in the interval $0<R_{0}<R$. Furthermore, from \eqref{interiorf} and \eqref{pressure} we notice that the pressure becomes negative in the regime $0\leq r < R_{0}$, but $e^{\nu}>0$. As the radius of the star approaches the Schwarzschild radius from above $R\to R_{S}^{+}$ meanwhile, $R_{0}\to R_{S}^{-}$ from below\footnote{This corresponds to a quasi-stationary adiabatic contraction.}. In the ultracompact limit when $R=R_{0}=R_{S}$, the Schwarzschild interior solution \eqref{interiorf} and \eqref{pressure} shows that the whole new interior region becomes one with constant negative pressure 

\begin{eqnarray}
p=-\rho,\quad r<R=R_{0}=R_{S}.
\end{eqnarray}

\begin{figure}
\label{fig1}
\centering
\includegraphics[scale=0.6]{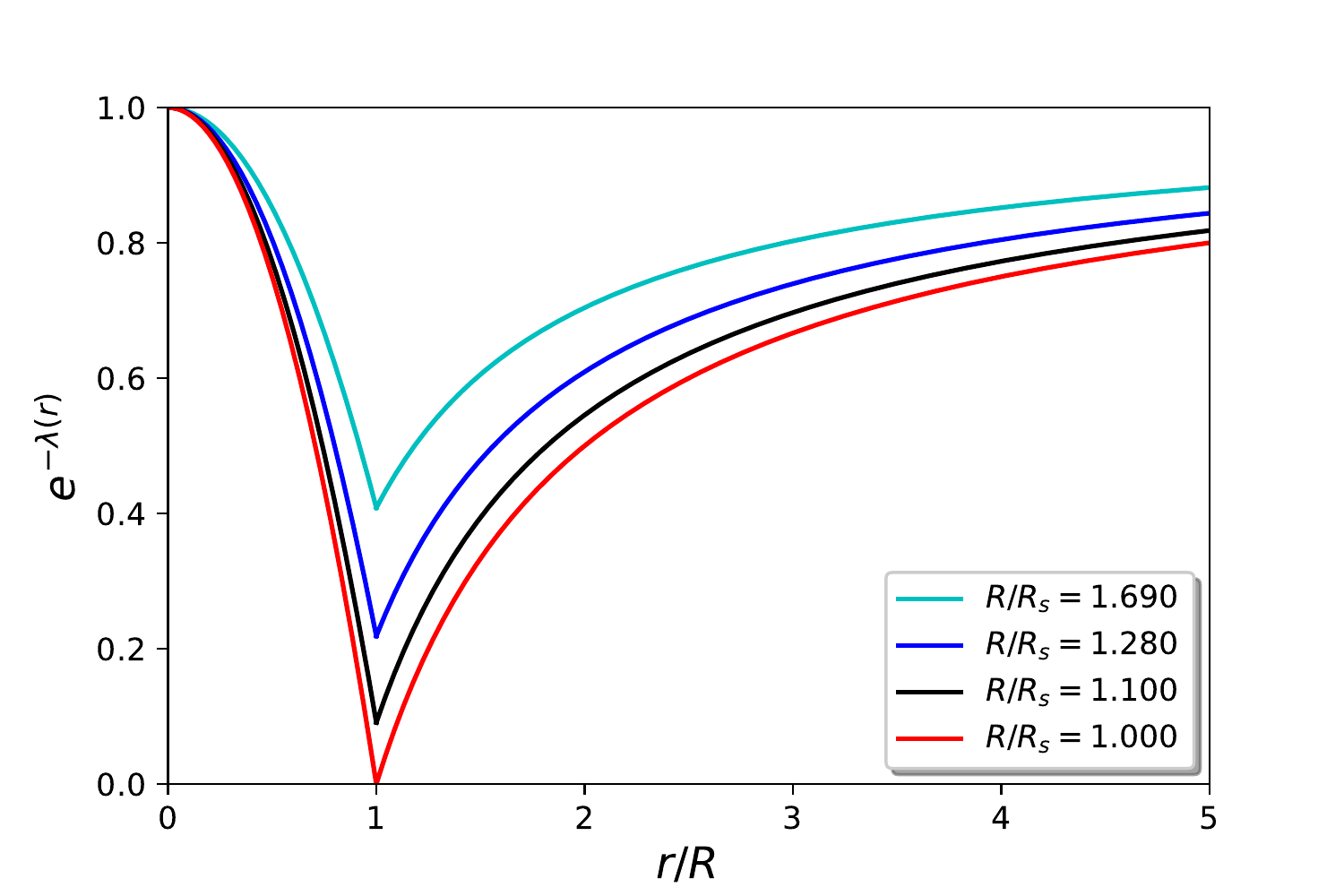}
\caption{\label{fig1} Metric function $e^{-\lambda(r)}$, as a function of $r/R$, for the interior and exterior of the Schwarzschild star. Note the ‘cusp-like' behaviour at  the matching point, and the approach of the minimum of $e^{-\lambda}$ to zero as the compactness approaches the ‘black hole' limit.}
\end{figure}

The interior metric function \eqref{interiorf} becomes a patch of de Sitter spacetime, and one where the $g_{tt}$ metric-component is modified by $1/4$ indicating that the time runs slower inside the configuration (see \cite{Mazur:2015kia} for further details). The final form of the interior metric is given by

\begin{eqnarray}\label{gravin}
e^{\nu(r)}=\frac{1}{4}(1-H^2r^2),\quad e^{-\lambda(r)}= (1-H^2r^2),\quad r<R_{0}=R_{S}.
\end{eqnarray}

\noindent Outside the configuration $r>R_{S}$, the spacetime geometry remains the spherically symmetric exterior Schwarzschild solution. Note from figure \ref{fig1} that the interior and exterior metric functions $g^{rr}$ are continuous at the surface $R=R_{0}=R_{s}$, but they join in a ‘cusp-like', non-analytic, behaviour which implies a violation of the second junction condition $[K_{ij}]=0$. From the general formalism of the Israel junction conditions \cite{Israel:1966rt,barrabes1991}, a violation of the second junction condition implies the presence of a $\delta$-distribution of stresses located on the hypersurface $R_{0}$, which is given by \cite{Mazur:2015kia} 

\begin{eqnarray}\label{delta}
(p_{\perp}-p)=\frac{1}{3}\frac{\rho R_{0}^3}{r^2}\left(\frac{h}{f}\right)^{1/2}\delta(r-R_{0}).
\end{eqnarray}

This assumption is crucial to provide a physical interpretation to the Schwrzschild star beyond the Buchdahl limit. Moreover, these transverse stresses produce a finite surface energy and a finite surface tension, which is proportional to the difference in surface gravities, and is given by

\begin{eqnarray}\label{tension}
\tau_{surf} = \frac{\Delta\kappa}{8\pi}=\frac{MR_{0}}{4\pi R^3}.
\end{eqnarray}

It is relevant to remark that the metric functions remain positive for $r=R_{0}\to R_{S}$, therefore there is no trapped surface in the interior $r<R_{S}$, and no emergence of an event horizon. This configuration has zero entropy and zero temperature, which indicates its nature as a condensate. Thus, in the ultracompact limit when $R=R_{0}=R_{S}$, the Schwarzschild star mirrors the main features of the non-singular gravitational condensate star, or gravastar, proposed in \cite{mazur2001,mazur2004}. In the next section we will extend this model to a more general scenario using the MGD-decoupling.  

\section{MGD-gravastars}\label{s4}
Let us start by identifying the gravastar solution~\eqref{gravin} with the undeformed metric~\eqref{pfmetric}. Hence the explicit form of the metric components $\{\xi,\,\mu\}$ reads

\begin{eqnarray}
\label{mm00}
&&e^{\xi}=\frac{1}{4}(1-H^2r^2)\ ,\ \\
\label{mm11}
&&e^{-\mu}=1-H^2r^2\ .
\end{eqnarray}

Next, we shall find the deformed version of the gravastar solution~\eqref{gravin} by using the MGD transformation~\eqref{expectg}. Hence the MGD-gravastar, generically represented by the MGD metric in~\eqref{mgdmetric}, is written as
\begin{eqnarray}
\label{grav00}
&&e^{\nu}=\frac{1}{4}(1-H^2r^2)\ ,\ \\
&&e^{-\lambda}=1-H^2r^2+\alpha\,f^{*}(r)\ ,
\label{grav11}
\end{eqnarray}
where the geometric deformation $f^{*}$ is found by solving the system~\eqref{ec1d}-\eqref{ec3d}. Since this system has four unknown functions, namely, the three components of the source $\theta_{\mu\nu}$ and the deformation $f^{*}$, we need to prescribe an additional condition. We may impose an equation of state for the source $\theta_{\mu\nu}$ or a physically motivated restriction for the deformation $f^{*}$. This last option will be the one we will choose next.
\par
A main characteristic of a gravastar is the  condition $g_{tt}=g^{-1}_{rr}=0$ at its surface $r=R_S$. In order to keep this critical condition in the deformed version, the geometric deformation $f^{*}$ should satisfy
\begin{eqnarray}
\label{fcondition}
f^{*}(r)\sim\,1-H^2r^2\ .
\end{eqnarray}
The simplest expression satisfying the requirement~\eqref{fcondition} is given by 
\begin{eqnarray}
\label{fcondition2}
f^{*}(r)=\left(1-H^2\,r^2\right)\,H^n\,r^n\ ,
\end{eqnarray}
where $n\geq\,2$ to avoid a singular solution [see also \eqref{GSefecden}-\eqref{GSanisotropy}]. Using the expression~\eqref{fcondition2}, the radial metric component~\eqref{grav11} reads
\begin{eqnarray}
\label{grav11d}
e^{-\lambda}=\left(1-H^2\,r^2\right)\left(1+\alpha\,H^n\,r^n\right)\ ,  
\end{eqnarray}
where 
\begin{eqnarray}
\label{alphacond}
\alpha\,\geq\,-1
\end{eqnarray}
to have a positive defined value in the expression~\eqref{grav11d} when $r\rightarrow\,R_S\,$.
\par
The MGD gravastar metric, which is a solution of Einstein's field equations~\eqref{ec1}-\eqref{ec3}, and whose explicit components are given by \eqref{grav00} and \eqref{grav11d}, generates an effective density 
\begin{eqnarray}
k^2\,\tilde{\rho}(r)=3\,H^2+\alpha\,H^n\,r^{n-2}\left[(n+3)H^2\,r^2-n-1\right]\ ,
\label{GSefecden}
\end{eqnarray}
an effective radial pressure
\begin{eqnarray}
k^2\,\tilde{p}_{r}(r)
=-3\,H^2-\alpha\,H^n\,r^{n-2}\left(3\,H^2\,r^2-1\right) 
\ ,
\label{GSefecprera}
\end{eqnarray}
and an effective tangential pressure
\begin{eqnarray}
k^2\,\tilde{p}_{t}(r)
=-3\,H^2-\alpha\,H^n\,r^{n-2}\left[(n+3)H^2\,r^2-\frac{n}{2}\right]\ .
\label{GSefecptan}
\end{eqnarray}
The anisotropy is given by
\begin{eqnarray}
\label{GSanisotropy}
\Pi
\equiv
\tilde{p}_{t}-\tilde{p}_{r}=\frac{\alpha\,H^n\,r^{n-2}}{2\,k^2}\left(n-2-2\,n\,H^2\,r^2\right)
\ .
\end{eqnarray}
We remark that the expressions in \eqref{grav00} and \eqref{grav11d}-\eqref{GSefecptan}, which describe a non-uniform and anisotropic ultracompact distribution (gravastar), satisfy Einstein's field equations \eqref{ec1}-\eqref{ec3}.
\par
By using the MGD gravastar solution displayed in~\eqref{grav00} and~\eqref{grav11d} in the first fundamental form~\eqref{ffgeneric1} and~\eqref{ffgeneric2}, we obtain 
\begin{eqnarray}
\label{ffgrav00}
&&0=1-\frac{2\,{\cal M}}{R}\ ,\\
&&0=1-\frac{2\,{\cal M}}{R}+\beta\,g^{*}_R
\label{ffgrav11}\ ,
\end{eqnarray}
which yields the important result
\begin{eqnarray}
\label{nodef}
g^{*}_R\sim\,1-\frac{2\,{\cal M}}{R}=0\ .
\end{eqnarray}
The condition~\eqref{nodef} is crucial since it establishes that the geometric deformation $g^{*}(r)$ must vanish at the stellar surface of a gravastar, no matter the nature of the exterior source $\theta_{\mu\nu}$. On the other hand, by using the effective radial pressure~\eqref{GSefecprera}, the continuity of the second fundamental form in \eqref{sfgenericf} yields
\begin{eqnarray}
\label{sffgrav}
-\left(3+2\,\alpha\right)=\frac{\beta\,g_{R}^{*}}{(R-2\,{\cal M})}
\ .
\end{eqnarray}
 The expressions in \eqref{ffgrav00}), \eqref{nodef} and \eqref{sffgrav} are the necessary
 and sufficient conditions for the matching of the interior metric \eqref{metric}, explicitly  displayed in \eqref{grav00} and \eqref{grav11d}, to a deformed exterior Schwarzschild metric in {\eqref{MetricSds}}.  We remark that conditions~(\ref{ffgrav00}) and (\ref{nodef}) do not necessarily imply a singular or null right hand side in Eq.~\eqref{sffgrav}. We notice that for a Schwarzschild exterior, namely $g^{*}(r)=0$, the expression~\eqref{sffgrav} yields $\alpha\,=-3/2$, which is incompatible with the regularity condition~\eqref{alphacond}. Therefore, we conclude that the MGD gravastar solution \eqref{grav00} and \eqref{grav11d} is ill-matched with the Schwarzschild exterior solution.
 \par 
  \begin{figure}[t]
  	\center
  	\includegraphics[scale=0.6]{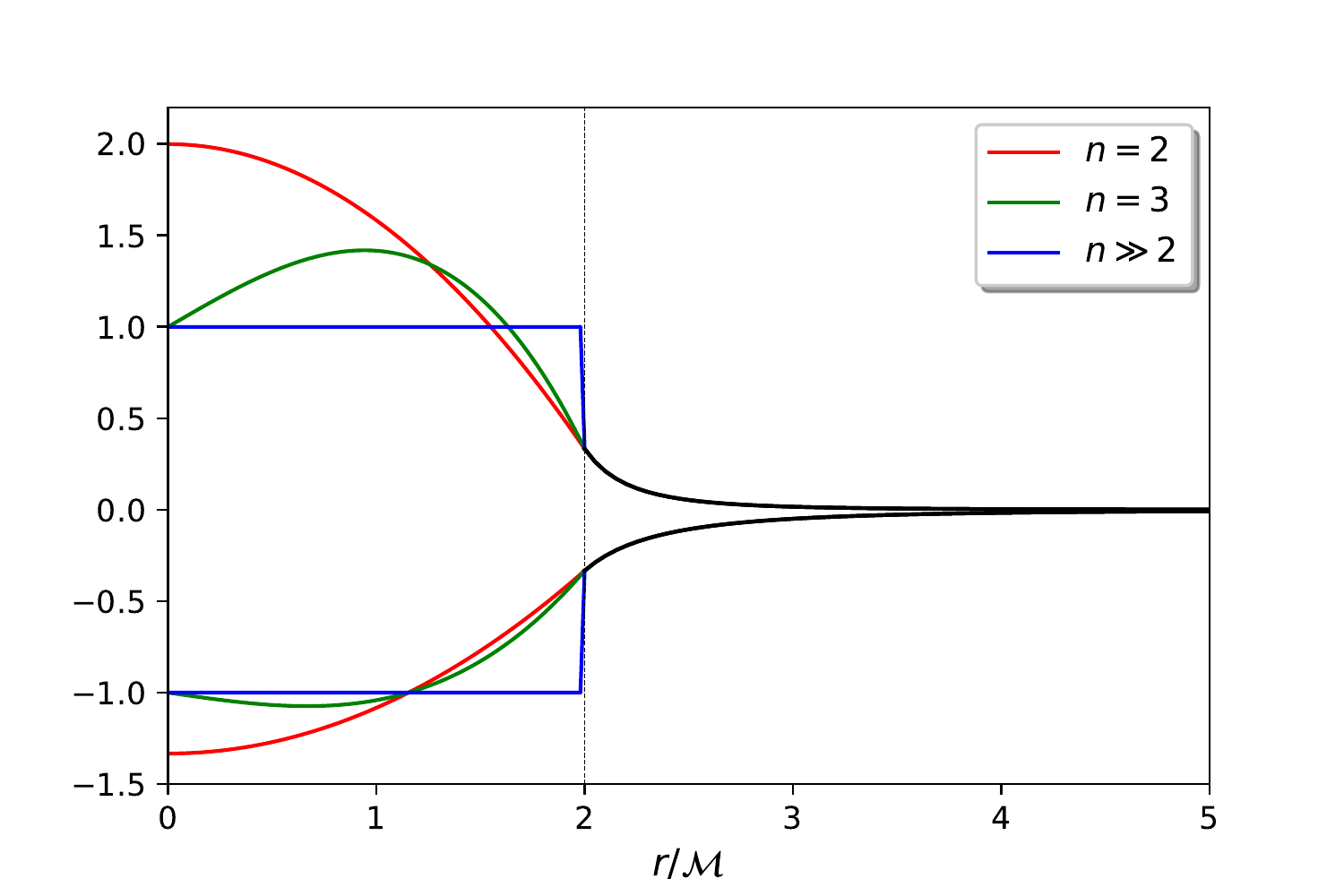}
  	\\
  	\caption{Behavior of the interior density $\tilde{\rho}>0$ and interior radial pressure $\tilde{p}_r<0$ for $n=2$, $n=3$ and $n>>2$.  The exterior $r>2\,{\cal M}$ is filled with a spherically symmetric source $\theta_{\mu\nu}$ which goes to zero quickly (black line). There are two stable configurations, namely, $n=2$ and the extreme case $n>>2$ (Mazur-Mottola model).}
  	\label{fig2}      
  \end{figure}
  \begin{figure}[t]
 	\center
 	\includegraphics[scale=0.6]{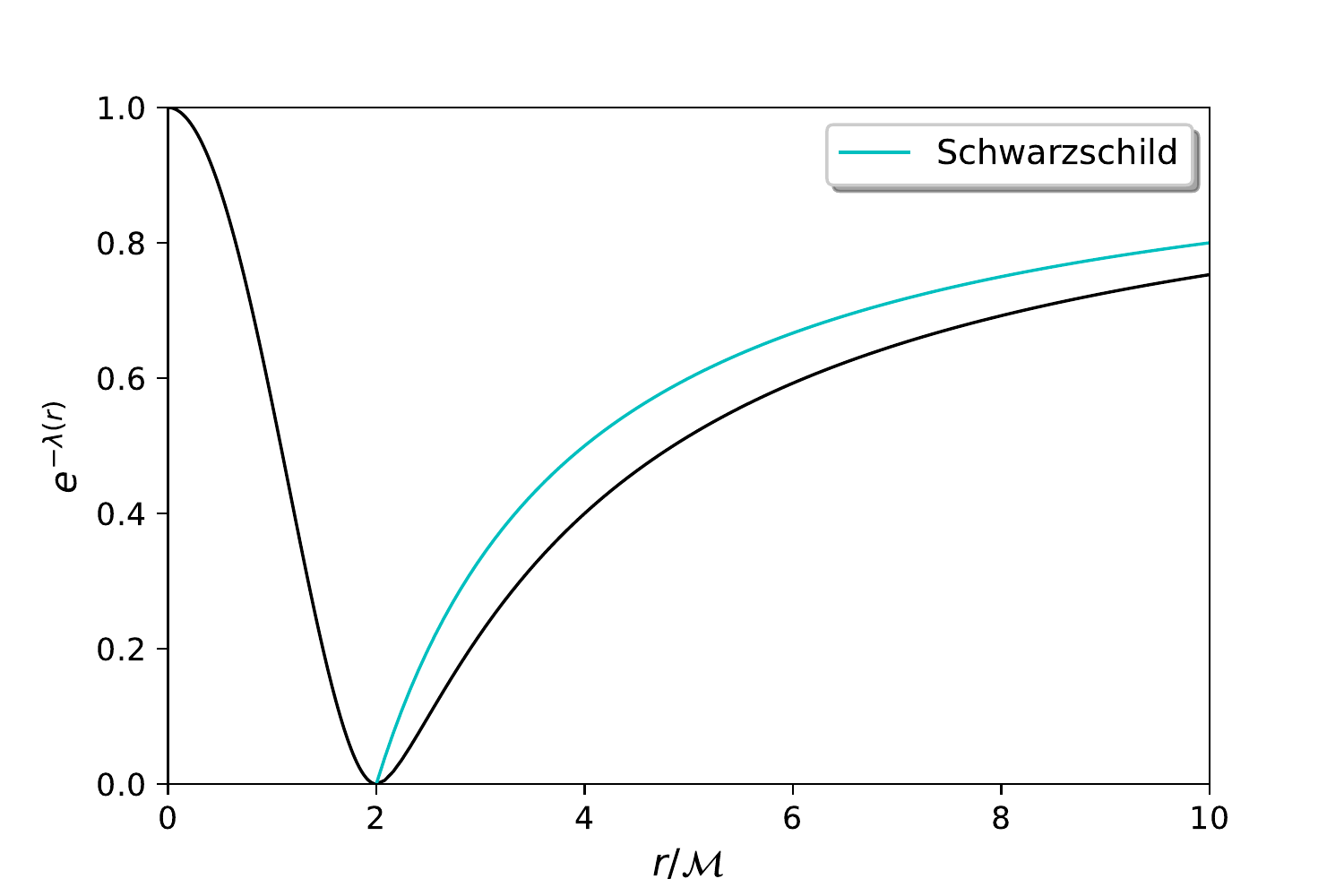}
 	\\
 	\caption{Interior and exterior radial metric component for the MGD-gravastar, as a function of $r/\cal{M}$. In contrast to the ‘cusp-like' behaviour in figure 1, note here the smooth continuity of the metric component at the stellar surface.}
 	\label{fig3}      
 \end{figure}
 \par
 Next, we will solve the  system~\eqref{ec1de}-\eqref{ec3de} to find an exterior solution $\{\theta^{+}_{\mu\nu},\,g^{*}\}$ compatible with the MGD gravastar in \eqref{grav00} and \eqref{grav11d}. First of all, we notice that the system \eqref{ec1de}-\eqref{ec3de} has four unknown functions. Hence, we need to provide additional information. We have two alternatives: either an equation of state associated with the source $\theta^{+}_{\mu\nu}$ or some physically motivated restriction on $g^{*}$. Among all possibilities, we choice an exterior source $\theta^{+}_{\mu\nu}$ associate with a conformal symmetry, hence
 
 \begin{eqnarray}
 \label{tra}
 \theta^{+\,\mu}_{\,\mu}=0\ .
 \end{eqnarray}
 
 We will see that the traceless condition~\eqref{tra} yields a conformally deformed exterior compatible with the interior solution displayed in \eqref{grav00} and \eqref{grav11d}. By using \eqref{ec1de}-\eqref{ec3de} in the condition \eqref{tra} we find that the radial deformation must satisfy the following differential equation

\begin{eqnarray}
\label{fconf}
r\,\left(6\,{\cal M}^2-7\,{\cal M}\,r+2\,r^2\right)\,g^{*'}+2\left(r^2-4\,{\cal M}\,r+3\,{\cal M}^2\right)\,g^{*}
=
0
\ ,
\end{eqnarray}
 whose general solution is given by
\begin{eqnarray}
\label{gconf}
g^{*}(r)
=
\frac{1-2\,{\cal M}/r}{2\,r-{3\,{\cal M}}}
\,\ell_{\rm c}
\ , 
\end{eqnarray}
with $\ell_{\rm c}$ a constant with units of a length. Thus the conformally deformed Schwarzschild exterior becomes~\cite{Casadio:2017sze}
\begin{eqnarray}
\label{confsol}
e^{-\lambda}
=
\left(1-\frac{2\,{\cal M}}{r}\right)
\left(1+\frac{\ell}{2\,r-{3\,{\cal M}}}\right)
\ ,
\end{eqnarray}
where $\ell=\beta\,\ell_{\rm c}$, and its behaviour for $r\gg M$ is given by
\begin{eqnarray}
e^{-\lambda}
\simeq
1
-
\frac{4\,{\cal M}-\ell}{2\,r}
\ .
\end{eqnarray}
A solution similar to that in \eqref{confsol} was found in the context of the extra-dimensional brane-world \cite{Germani:2001du} and subsequently analyzed in detail in the context of MGD-black holes \cite{Ovalle:2017wqi}. By using \eqref{ec1de}-\eqref{ec3de} we compute the exterior physical variables, namely, the effective density

\begin{eqnarray}
\tilde{\rho}^+
=
\beta\,(\theta_0^{\ 0})^+
=
-\frac{\ell\,{\cal M}}{k^2\,(2\,r-3\,{\cal M})^2\,r^2}
\ ,
\label{efecdenC}
\end{eqnarray}
the effective radial pressure
\begin{eqnarray}
\tilde{p}^{+}_{r}
=
-\beta\,(\theta_1^{\ 1})^+
=
\frac{\ell}{k^2\,(2\,r-3\,{\cal M})\,r^2}
\ ,
\label{efecpreraC}
\end{eqnarray}
and the effective tangential pressure
\begin{eqnarray}
\tilde{p}^{+}_{t}
=
-\beta\,(\theta_2^{\ 2})^+
=
\frac{\ell\,(r-{\cal M})}{k^2\,(2\,r-3\,{\cal M})^2\,r^2}
\ .
\label{efecptanC}
\end{eqnarray}
The exterior anisotropy is thus given by
\begin{eqnarray}
\Pi^{+}
=
\frac{\ell\,(3\,r-4\,{\cal M})}{k^2\,(2\,r-3\,{\cal M})^2\,r^2}
\ .
\end{eqnarray}
\par
Finally, we see that the exterior deformation~\eqref{gconf} satisfies the matching conditions in Eqs.~(\ref{ffgrav00}) and~(\ref{nodef}), while the second fundamental form~(\ref{sffgrav}) yields
\begin{eqnarray}
\label{l}
\ell=-(3+2\,\alpha)\,{\cal M}\ .
\end{eqnarray}
Using the matching condition~\eqref{l} in the expression~\eqref{confsol}, we find the deformed radial metric component will always be positive for the region
\begin{eqnarray}
\label{defpos}
r\,\geq\,(3+\alpha)\,{\cal M}\ ,
\end{eqnarray}
therefore the exterior space-time will be regular for $\alpha$ satisfying 
\begin{eqnarray}
\label{alpha}
\alpha=-1\ .
\end{eqnarray}
We conclude the conformally deformed Schwarzschild exterior~\eqref{confsol} is consistent with the interior ultracompact configuration in \eqref{grav00} and \eqref{grav11d} for $\{\alpha=-1,\,\ell=-\,{\cal M}\}$ and any value of $n\geq\,2$ (notice that the left-hand side in the matching condition~\eqref{sffgrav} is independent of $n$). This is very significant since it indicates that the complete family of interior solutions, characterized by the parameter $n$, can be matched with the exterior solution. However, as we see in figure \ref{fig2}, only the case $n=2$ and the extreme case $n>>2$ (Mazur-Mottola model) are stables, namely, there is no maximum (or minimum) between $0\le\,r\le\,2\,{\cal M}$.

Another point which calls our attention is the continuity of the effective density $\tilde{\cal \rho}$ through $R_S$. The continuity of the effective radial pressure $\tilde{p}_r$ is a direct consequence of the second fundamental form, but in principle there is no reason for the density to be continuous through $R_S$. We can explain this by examining the derivative of the metric function $e^{-\lambda}$ near the stellar surface. Using the interior metric function~\eqref{grav11d}, we have
\begin{eqnarray}
\label{dg11i}
-\lambda'\,e^{-\lambda}\bigg\vert_{R_S^{-}} = -\frac{2}{R_S}\left(1+\alpha\right)\ ,
\end{eqnarray}
whereas the exterior metric function~\eqref{confsol} yields
\begin{eqnarray}
\label{dg11e}
-\lambda'\,e^{-\lambda}\bigg\vert_{R_S^{+}} = \frac{1}{R_S}\left(1+\frac{\ell}{\cal M}\right)\ .
\end{eqnarray}
Using the matching condition~\eqref{l} in Eq.~\eqref{dg11e}, we see that both expressions in~\eqref{dg11i} and~\eqref{dg11e} are equal. Hence the metric function $e^{-\lambda\,}$ is smoothly continuous through the stellar surface, as we can see in Figure~\ref{fig2}. Also the condition~\eqref{alpha} states that the function $e^{-\lambda\,}$ is always positive. 
Finally, we see that the continuity of $(e^{-\lambda})'$ yields a continuous effective density $\tilde{\cal \rho}$, as we can see through the field equation~\eqref{ec1}.
\par
\begin{figure}[t]
	\center
	\includegraphics[scale=0.6]{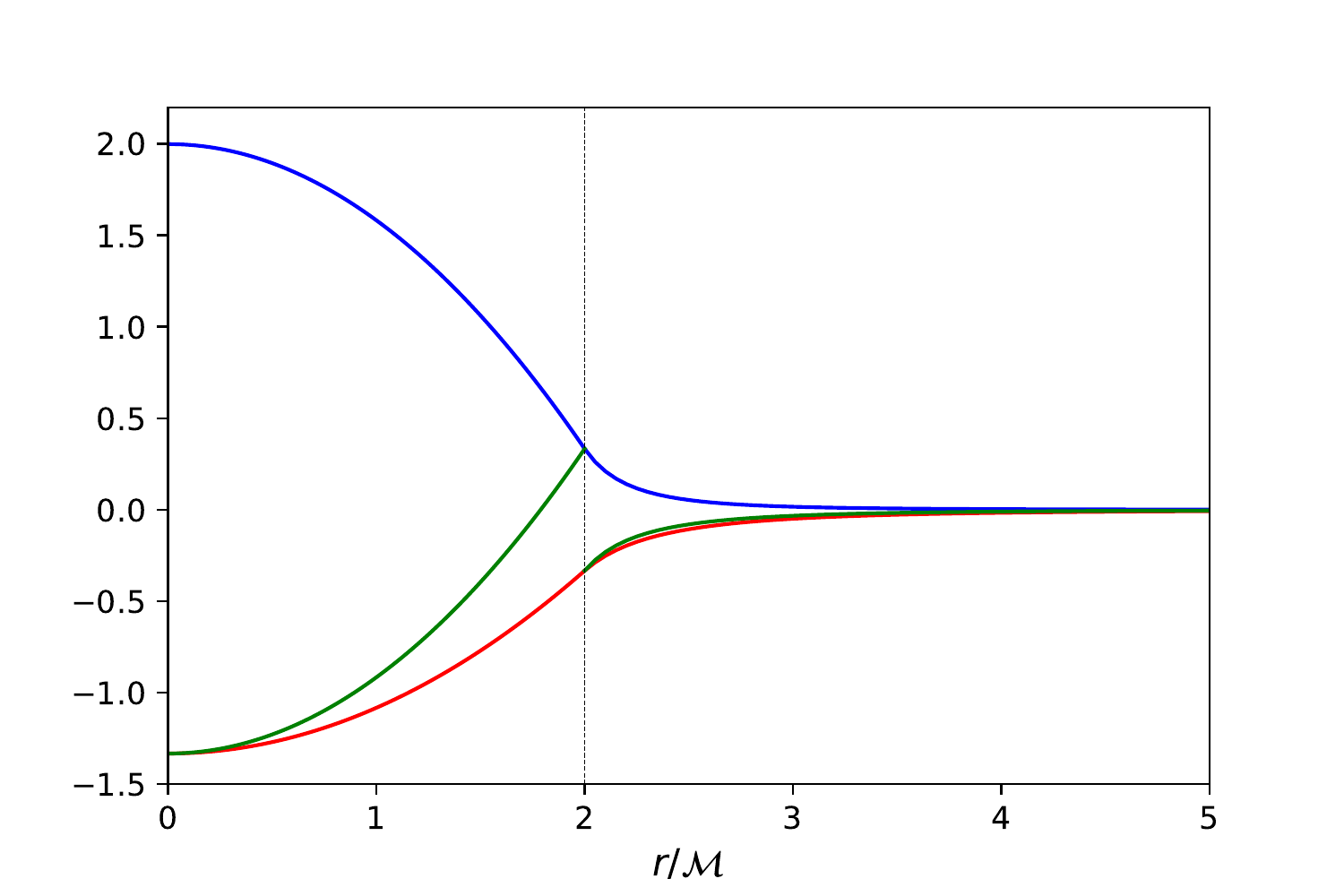}
	\\
	\caption{Case $n=2$. Behavior of the density $\tilde{\rho}$ (blue line), radial pressure $\tilde{p}$ (red line) and tangential pressure $\tilde{p}_t$ (green line). The discontinuity in the  tangential pressure occurs at $r=R_{S}=2\,{\cal M}$.}
	\label{figure4}      
\end{figure}
Additionally, it is worth to analyse the continuity of the radial pressure accross the surface of the configuration. Expressions for the interior and exterior radial pressure in~\eqref{GSefecprera} and~\eqref{efecpreraC}, respectively, yield
\begin{eqnarray}
\label{dpi}
&&\frac{d\,{\tilde p}_r}{d\,r}\bigg\vert _{R_S^{-}} = -\frac{2\,\alpha}{k^2 R_S^3}\,(n+1)\ ,
\\
\label{dpe}
&&\frac{d\,{\tilde p}_r}{d\,r}\bigg\vert_{R_S^{+}} = \frac{6}{k^2\,R^2_S}\ .
\end{eqnarray}
We see that precisely the case $n=2$, together with the regularity condition~\eqref{alpha}, yields a radial pressure which is smoothly continuous through the  stellar surface. This remarkable feature suggests that the stellar system is extended beyond its defined surface $r=R_{S}$. However, the discontinuity in the tangential pressure at $r=R_{S}$, as we can see in figure \ref{figure4}, clearly establishes the surface of the self-gravitating distribution. This condition is reminiscent of the boundary layer of anisotropic stresses which was introduced for the ultracompact Schwarzschild star, discussed in section \ref{s3}.   
\par
We conclude this section by pointing out that the dominant energy condition
\begin{eqnarray}
\tilde{\rho}\,\geq\,\mid\,\tilde{p}_r\mid\ ;\,\,\,\,\,\tilde{\rho}\,\geq\,\mid\,\tilde{p}_t\mid\ ,
\end{eqnarray}
is satisfied inside the ultracompact configuration.
\section{Conclusions}
\label{con}
\setcounter{equation}{0}
By using the MGD-decoupling approach, we propose an anisotropic and non-uniform version of the gravastar model by Mazur and Mottola, given by the exact and analytical expressions \eqref{grav00}, \eqref{grav11d} and \eqref{GSefecden}-\eqref{GSefecptan}. This new system represents an ultracompact configuration of radius $R_S=2\,{\cal M}$ which satisfies some of the requirements to describe a stable stellar model, namely, it is regular at the origin, its mass and radius are well defined, its density is positive everywhere and decreases monotonically from the centre outwards, as well as a non-uniform and monotonic pressure. Additionally this new solution satisfies the dominant energy condition. These features show that the anisotropic effects produce a more realistic stellar structure, but above all, it indicates the possibility of building ultracompact configurations of radius $R_S=2\,{\cal M}$, minimising their exotic characteristics.
\par
A crucial point in the construction of the new solution is to preserve the null-surface condition $g_{tt}=g^{-1}_{rr}=0$ at $R_S=2\,{\cal M}$, which is a fundamental property of the ultracompact Schwarzschild stars, or gravastars. This yields the metric component~\eqref{grav11d} characterized by a parameter $n\ge\,2$, where $n>>2$ represents the Mazur-Mottola model. On the other hand, the exterior space-time is represented by a deformed Schwarzschild solution~\eqref{MetricSds}, whose deformation $g^{*}$ is produced by a generic conserved energy-momentum tensor $\theta^{+}_{\mu\nu}$ filling the Schwarzschild vacuum  $T^{+}_{\mu\nu}=0$. Hence the exterior is a ‘tensor-vacuum' $\{T^{+}_{\mu\nu}=0,\,\theta^{+}_{\mu\nu}\,\neq\,0\}\,$\cite{Ovalle:2017fgl,Ovalle:2019qyi}.
\par
 In contrast to most of the gravastar models, in our case we avoid the introduction of thin-shells of matter or any other mechanism beyond the simple and well-established Darmois-Israel matching conditions at the surface. In particular, the continuity of the metric implies that the deformation $g^{*}$ of the Schwarzschild exterior must vanish at the stellar surface for any ultracompact configuration of radius $R_S=2\,{\cal M}$. This result is independent of the nature of $\theta^{+}_{\mu\nu}$, as it is established in the condition~\eqref{nodef}.
\par
Among all the possibilities, we chose a conformally deformed vacuum defined by \eqref{tra}, which yields the deformed Schwarzschild exterior \eqref{confsol}. The main characteristic of this solution is that for $r>>{\cal M}; \,$ $\tilde{\rho}\sim\,r^{-4},\,$ $\,
\tilde{p}_r\sim\tilde{p}_t\sim\,r^{-3}$. Hence the physical variables decay quickly. In particular, the continuity of the second fundamental form in \eqref{l} establishes that the conformally deformed Schwarzschild exterior \eqref{confsol} can be matched to the interior solution \eqref{grav00} and \eqref{grav11d} for any value of $n\geq\,2$, and that not only the radial pressure $\tilde{p}_r$ but also the density $\tilde{\rho}$ is always continuous accross $R_S$. However,  only the case $n=2$ and the extreme case $n>>2$ (Mazur-Mottola model) are stables, as we see in figure \ref{fig2}. 

We also point out that the metric function $e^{-\lambda\,}$ is smoothly continuous through the stellar surface for any value of $n$, as we can see in figure \ref{fig2}. Therefore the effective density $\tilde{\cal \rho}$ will always be continuous as a consequence of the field equation~\eqref{ec1}. Finally the case $n=2$, which is the physically relevant in terms of stability, along with the regularity condition~\eqref{alpha}, yields a radial pressure smoothly continuous through the  stellar surface as figure \ref{figure4} shows.
\par
Finally, we conclude by rising some natural questions regarding the analysis of ultracompact configurations under the MGD-decoupling which deserve to be investigated, such as: the analysis of stability under spherical and non-spherical perturbations; any other physically motivated ``tensor-vacuum" $\{T^{+}_{\mu\nu}=0,\,\theta^{+}_{\mu\nu}\,\neq\,0\}$ generating reasonable consequences; a complete deformation of the gravastar solution in Eqs.~\eqref{mm00} and~\eqref{mm11} by the extended deformation~\cite{Ovalle:2017fgl} in Eqs.~\eqref{gd1} and~\eqref{gd2}, and a possible extension for time-dependent configurations.  

\section{Acknowledgements}
J.O.~and S.Z.~have been supported by the Albert Einstein Centre for Gravitation and Astrophysics financed
by the Czech Science Agency Grant No.14-37086G and by the Silesian University in Opava internal
grant SGS/14/2016. C. P. acknowledges the support of the Institute of Physics and Research Centre of Theoretical Physics and Astrophysics, at the Silesian University in Opava. 

%
\bibliography{references}
\bibliographystyle{iopart-num.bst}
\end{document}